\documentstyle[aps,prl,multicol,epsf]{revtex}

\newcommand{\bleq}{\ifpreprintsty
                   \else
                   \end{multicols}\vspace*{-3.5ex}{\tiny
                   \noindent\begin{tabular}[t]{c|}
                   \parbox{0.493\hsize}{~} \\ \hline \end{tabular}}
                   \fi}
\newcommand{\eleq}{\ifpreprintsty
                   \else
                   {\tiny\hspace*{\fill}\begin{tabular}[t]{|c}\hline
                    \parbox{0.49\hsize}{~} \\
                    \end{tabular}}\vspace*{-2.5ex}\begin{multicols}{2}
                    \fi}
\newcommand{\bcols}{\ifpreprintsty\else\begin{multicols}{2}\fi}
\newcommand{\ecols}{\ifpreprintsty\else\end{multicols}\fi}

\begin{document}
\bibliographystyle{prsty}
\title{Weak Charge Quantization as an Instanton of
 Interacting $\sigma$--model}  
\draft
\author{ Alex Kamenev}
\address{ Department of Physics, Technion, Haifa 32000, Israel.
  \\
  {}~{\rm (\today)}~
  \medskip \\
  \parbox{14cm} 
    {\rm Coulomb blockade in a quantum dot attached to a  
    diffusive conductor is considered in the framework of the  
    non--linear $\sigma$--model. It is shown that the weak charge 
    quantization on the dot 
    is associated with   instanton configurations of the 
    $Q$--field in the conductor. 
    The  instantons have a finite action and are replica 
    non--symmetric.  It is argued that such instantons may play a   
    role in the transition regime to the interacting insulator. 
    }\bigskip \\ }
\maketitle
\bcols
The physics of interacting electronic systems in the 
presence of disorder  has been 
a subject of an intense study  already for a few decades 
\cite{book}. 
Various theoretical approaches have been 
developed for description of both metallic and insulator  phases. 
The non linear $\sigma$--model (NL$\sigma$M) in the replica 
\cite{Finkelstein90,Belitz94}
or dynamic 
\cite{Kamenev99,Chamon99}
formulation has proven to be the most powerful tool to deal with the weakly 
disordered (metallic) phase. Despite of a lot of efforts, 
the NL$\sigma$M was not really successful in treating the insulating 
and transition regimes. On the  other hand,    
approaches based on the idea of hopping conductivity
\cite{Efros85} 
are quite efficient in the phenomenological description of the 
insulating phase.
This is not the case in the theory of non--interacting
localization, where the NL$\sigma$M provides a unified treatment \cite{Efetov} of
both metallic and insulating regimes (at least in low dimensions, 
$d\leq 2+\epsilon$).   It is also known to give some insight 
\cite{Pruisken84}
into the physics of the integer quantum Hall transitions. 

To describe the insulating phase one usually assumes that the system may be 
divided into weakly connected islands whose charge is essentially quantized.  
The charge may be changed only by an addition or subtraction of an  
integer amount of $e$, which costs a  finite Coulomb energy. 
As a result, the important 
charge configurations are quantized and  inhomogeneous. 
Approaching the metallic phase, the coupling between islands becomes
better and charge quantization (Coulomb blockade) effects are less 
pronounced. Finally, deep in the metallic phase the charge is continuous and 
practically  homogeneous (due to screening). In a {\em perturbative} 
treatment of  NL$\sigma$M (which is usually involved in description of the 
metallic phase) there is no room for the discreetness of the electron charge.
It is doubtful,  therefore, that  any NL$\sigma$M perturbative approach can  
provide a picture of the transition or insulating regimes, where 
charge quantization is of primary importance. 
The purpose of this Letter is to show how the weak residual 
charge quantization
may be incorporated  in the  diffusive NL$\sigma$M formalism.

To this end I consider a toy problem of a large quantum dot with 
the capacitive interaction connected to a bulk reservoir by a diffusive 
non--interacting wire. 
The dot is open in a sense that the wire conductance 
is large: $G \gg G_Q \equiv e^2/(2 \pi \hbar)$. One may, nevertheless, ask
whether any Coulomb blockade effects can be seen on such a dot. 
In particular, we shall look for the oscillatory part of the 
electron number, $\tilde N(q)$, on the dot as a function of the continuous
background charge, $q$. (For a closed dot, $G\ll 1$, such  
oscillations are, of course, extremely pronounced  
\cite{Likharev}.) In a limiting case of a short (see below) wire this problem 
was recently solved by Nazarov \cite{Nazarov99}. 
Employing the model introduced earlier by Matveev \cite{Matveev95}, 
he derived an expression for an amplitude, $A$, of the Coulomb blockade 
oscillations for a  dot connected to a reservoir by an 
arbitrary (delay free) scattering region:
\begin{equation} 
A \sim \prod\limits_{i} |R_i|^{1/2}\, ,
                                                             \label{Naz}
\end{equation}
where $R_i$ are  eigenvalues of   the scattering region reflection matrix. 
Then, e.g. 
$\tilde N(q) = A\cos(2\pi q +\eta)$,  
where $\eta$ is a phase, which depends on the scattering matrix of the contact. 
For a diffusive wire the eigenvalues $R_i$ are  random 
functions of the  disorder realization with the probability density
\cite{Nazarov94} 
$P(R) = G/[2(1-R)\sqrt{R}]$, 
where $G=\int dR (1-R) P(R)$ is the average conductance of the wire
(hereafter conductance is measured in units of $G_Q$). Employing 
Eq.~(\ref{Naz}), one finds for the typical oscillation amplitude 
\begin{equation} 
e^{\langle \ln A \rangle} = e^{-{\pi^2\over 8}G }\, ,                  
                                            \label{Naz1}
\end{equation}
where the angular brackets stand for  disorder averaging. This result 
holds for a metallic wire with $L \ll \xi$, where $L$ is the wire length
and 
$\xi$ is localization length, 
and therefore $G\gg 1$. Moreover, to treat the wire as an instantaneous 
scattering region one assumes
that 
$L \ll L_T\equiv \sqrt{ \hbar D/(2\pi T)}$ \cite{foot1}, where $D$ is  
diffusion constant of the wire and $T$ is  temperature. 
Since, according to Eq.~(\ref{Naz1}), the Coulomb blockade is exponentially 
suppressed, we refer to it as the weak charge quantization.

We are  interested  in the wire rather than in the dot. One would
like  to show, thus, how the weak charge quantization, 
Eqs. (\ref{Naz}), (\ref{Naz1}), may be understood (and extended for $L>L_T$) 
from the point of view  of the NL$\sigma$M of the wire. 
Since the coupling constant of the NL$\sigma$M is $1/G$, the expected result,
Eq.~(\ref{Naz1}),  is the non--analytic function of the coupling constant. 
Therefore, it can not be obtained in any  perturbative treatment 
of NL$\sigma$M. Indeed, we shall show that these are  the instanton 
configurations 
with the finite action, which are responsible for  the weak charge quantization. 
In our toy model  the electron--electron interactions and disorder
are spatially separated on the dot  and the wire correspondingly. 
Nevertheless,  
one has to consider the full interacting NL$\sigma$M of Finkel'stein 
\cite{Finkelstein90,Belitz94} 
with the $Q$--field having a structure both in the energy and replica spaces. 
The finite action instantons, mentioned above, involve  rotations between two 
or more different replicas. Thus each instanton configuration is {\em replica 
non--symmetric}. Eventually the replica symmetry is restored by  instantons 
with all possible replica permutations. 
The model provides  the first  example 
of a non--perturbative and non replica--diagonal application
 of the interacting NL$\sigma$M.
We  confirm Nazarov's result, 
Eq.~(\ref{Naz1}),  and  extend it to the case of 
a long, $L > L_T$, wire and 2D diffusive conductor.

Let us consider a large \cite{foot2} quantum dot connected to 
a reservoir by a non-interacting diffusive wire. 
The Coulomb interaction on the dot has the form 
\begin{equation} 
H_{int} = E_c (\hat N - q)^2\, ,
                                         \label{Ham}
\end{equation}
where $E_c=e^2/(2C)$ is the capacitive charging energy, 
$\hat N$ is the electron number operator on the dot and $q$ is the 
background charge. Our aim is to calculate the oscillatory part of the average 
(over quantum fluctuations) number of electrons on the dot, 
$\tilde N(q)$. The next step is to average the result over realizations of 
disorder in the diffusive wire. The random phase of  oscillations 
results in 
$\langle \tilde N(q) \rangle = 0$. Therefore, to calculate a typical 
oscillation amplitude one has to look for the correlation function 
\cite{foot4}
\begin{equation} 
\langle \tilde N(q) \tilde N(q') \rangle =
\lim_{n\to 0} {1\over n^2} 
\partial_{q}\partial_{q'} 
\langle Z^n(q) Z^n(q') \rangle\, ,
                                         \label{corr}
\end{equation}
where we introduced the partition function of the dot--wire system, $Z(q)$, along with 
the replica trick 
to facilitate the disorder averaging. 
The partition function of the open dot may be written in a standard way 
\cite{Schoen90}
as the imaginary time functional  integral over the scalar field, $\Phi(\tau)$. 
If $T\gg \Delta$ \cite{ffot2}, 
the $\Phi$ field configurations may be classified 
by the integer winding number, $W$, by imposing the boundary condition 
$\Phi(\tau + \beta) = \Phi(\tau) + 2\pi W$, where 
$\beta =1/T$. In our case both the field and the corresponding winding 
number are  $2n$--component replica--vectors, 
${\bf \Phi}=\{ \Phi_1,\ldots \Phi_{2n} \}$ and 
${\bf W}=\{ W_1,\ldots W_{2n} \}$. The correlation
function takes the form
\begin{eqnarray} 
\langle Z^n(q) &&Z^n(q') \rangle\! = \!\!\!
\sum\limits_{{\bf W} = -\infty}^{\infty}\!\!\! 
\exp\left\{ 
2\pi i\! \left(\!
q\!\sum\limits_{a=1}^n W_a\! + \!
q'\!\!\!\sum\limits_{a=n+1}^{2n}\! W_a\!
\right)\! 
\right\} \nonumber \\
\times &&\int\!\! {\cal D}{\bf \Phi}\, 
\exp\left\{\!\!
-{1\over 2E_c} \int\limits_0^\beta \!\! 
\sum\limits_{a=1}^{2n} \dot\Phi_a^2 d\tau - 
S_{w}({\bf \Phi}) \right\}\, . 
                                         \label{int}
\end{eqnarray}
Here $ S_{w}({\bf \Phi})$ is the average effective action which is a  result 
of tracing out degrees of freedom of the wire. The latter are connected to the dot's 
variable, ${\bf \Phi}$, through the boundary condition at the 
dot--wire interface (for $T\gg \Delta$). 
In terms of the NL$\sigma$M 
\cite{Finkelstein90,Belitz94}
the effective action may be written as
\begin{equation} 
e^{- S_{w}({\bf \Phi}) }\! = \!\!\!
\int\!\! {\cal D} {\bf Q} \,
\exp\! \left\{ \! \!
{\pi\nu\over 4}  \!\! \int\limits_0^L\!\!\! dx\,  
\mbox{Tr}\{
- D(\nabla {\bf Q})^2 \! + 
\! 4{\bf \hat \varepsilon Q} \} \! \! \right\} ,
                                            \label{nlsm}
\end{equation}
where $\nu$ is the density of states (DOS) in the wire and 
$ {\bf \hat \varepsilon }=\delta^{ab} \varepsilon_m \delta_{mk}$   
is a matrix diagonal in replica and Matsubara spaces, 
$\epsilon_m=2\pi T(m+{1\over 2})$. 
The ${\bf Q}$--field is also a matrix, 
$Q^{ab}_{\varepsilon_m\varepsilon_k}(x)$, 
both in the replica space, $a,b =1, \ldots 2n$  and the Matsubara space, 
$m,k = 0,\pm 1,\pm 2, \ldots$ 

On the wire--reservoir and the
wire--dot 
interfaces  the ${\bf Q}$--field obeys the following boundary conditions: 
\begin{mathletters}
\label{bc}
\begin{eqnarray} 
&& {\bf Q}(x=L) = {\bf \Lambda}
\equiv \delta^{ab}\mbox{sign}(\varepsilon_m)\delta_{mk}\, ;
                                            \label{bc1} \\
&& {\bf Q}(x=0) = 
e^{-i{\bf\Phi}(\tau)} {\bf \Lambda}_{\tau -\tau'} 
e^{i{\bf\Phi}(\tau')}\, ,  
                                            \label{bc2}
\end{eqnarray}
\end{mathletters} 
correspondingly. 
In the last equation we employed imaginary time representation of the ${\bf Q}$--field 
instead of the frequency representation. In addition, the normalization condition  
${\bf Q}^2 = 1$ must be obeyed at every point in  space. 

For a good metal, $G\equiv h\nu D/L \gg 1$, 
the ${\bf Q}$--integral may be calculated in the saddle point approximation. 
The extremal ${\bf Q}$ configurations are given by  the Usadel equation   
\begin{equation} 
\nabla(D{\bf Q}\nabla {\bf Q} ) -  
[{\bf \hat \varepsilon}, {\bf Q}] =0\, ,
                                          \label{usad}
\end{equation}
where {\bf Q} satisfies ${\bf Q}^2=1$ and the boundary 
conditions~(\ref{bc}). Our strategy, thus,  is to find the solution  
of Eq.~(\ref{usad}) for a fixed ${\bf \Phi}$ and identify 
$S_{w}({\bf \Phi})$ with  the action on 
the extremal ${\bf Q}$ configuration.

The oscillatory (in $q-q'$) component of the correlation function originates from 
non--zero winding numbers, cf. Eq.~(\ref{int}). 
Before proceeding in this direction, we shall make a few remarks about the 
${\bf W} =0$ sector. 
In this case the functional integral on the r.h.s. of  Eq.~(\ref{nlsm}), may be 
minimized  by the {\em replica diagonal} anzatz
\cite{Kamenev99}
\begin{equation}
{\bf Q}_{\tau\tau'}(x)= e^{-i{\bf K}(\tau,x)}
{\bf \Lambda}_{\tau-\tau'} 
e^{i{\bf K}(\tau',x)}\, ,
                                    \label{rdiag}
\end{equation}
where ${\bf K} =K_a(\tau,x)$ is replica and imaginary time diagonal matrix   
with  ${\bf K}(\tau,L)=0$ and ${\bf K}(\tau,0)={\bf \Phi}(\tau)$.
Substituting the anzatz (\ref{rdiag}) into Eq.~(\ref{nlsm}) and 
minimizing the action  with respect to ${\bf K}$ \cite{foot3},  one finds 
\begin{equation}
S_{w}({\bf \Phi})\! = \! {G\over 4\pi \beta}
\sum\limits_{\omega_m} 
{|\omega_m|^{3/2} \over E_T^{1/2} }
\coth \sqrt{ {|\omega_m| \over E_T}} \, 
|{\bf \Phi}(\omega_m)|^2 \, , 
                                       \label{Swire}
\end{equation}
where $E_T\equiv \hbar D/L^2$ is the  Thouless energy of the wire. In the short wire
limit, $|\omega_m|\ll E_T$, one recovers the phenomenological 
action of an ohmic environment
\cite{Leggett83,Schoen90} 
$S_{w}=G/(4\pi\beta) \sum_m |\omega_m||{\bf \Phi}_m|^2$.  
Although configurations with ${\bf W}=0$ do not affect the thermodynamics
of the dot, cf. Eq.~(\ref{int}), they change  its tunneling  DOS 
$\sim \langle \exp \{i({\bf \Phi}(\tau) -
{\bf \Phi}(\tau') \} \rangle_{\bf \Phi}$. 
In the small temperature   (or short wire) limit,  $T\ll E_T$, one finds the well--known 
\cite{Devoret90}
power--law    DOS $\sim T^{2/G}$.  
In the other limiting case, $T\gg E_T$,  
the DOS is given by $ \exp\{-{2\over G} \sqrt{{E_T\over T} }\} $.
This is nothing but the 1D Altshuler-Aronov  
zero-bias anomaly \cite{Altshuler85,Levitov97,Kamenev99} 
for the case of interacting dot and non--interacting wire.
One can easily include long--range  interactions inside the wire, 
which  lead to the dynamical screening term in the $K$ action 
\cite{Kamenev99,foot3}, to obtain the standard 1D result  
\cite{Altshuler85}. 
Therefore, both the phenomenological theory of an ohmic bath and the
zero--bias anomaly are   consequences
of the ${\bf W}=0$,  replica--diagonal sector of the NL$\sigma$M.

We turn now to   our main subject --- the non--trivial winding numbers,  
${\bf  W}\neq 0$. 
One may try to solve Eq.~(\ref{usad}) by the replica diagonal
anzatz, Eq.~(\ref{rdiag}). 
Then  the  phase  
$K_a(\tau,x)$ acquires $W_a$  vortices in the $(x,\tau)$ plane. 
As a result,  the  corresponding action is   logarithmically large,  
$S_{w} \sim G \ln L/\lambda$,  
where $\lambda$ is a microscopic cutoff length. 
(Note that this logarithmic divergence continues well in 
the ballistic regime, therefore $\lambda$ is likely to be of 
the order of the Fermi wavelength.) 
On the other hand, one may find  solutions of 
Eq.~(\ref{usad}) with much smaller  action, $S_{w} \sim G$. 
For  this one {\em must} allow   rotations between different   
replicas and impose the following  condition on the possible 
${\bf W}$ configurations: 
\begin{equation} 
\sum\limits_{a=1}^{2n} W_a = 0\, .
                                        \label{rest}
\end{equation}
Equations~(\ref{corr}), (\ref{int}) and  (\ref{rest})  immediately lead
to the following 
conclusions: (i) $\langle \tilde  N(q) \rangle = 0$; 
(ii) the correlation function  
$\langle \tilde N(q) \tilde N(q')  \rangle$ is a function of $q-q'$ only. 

To demonstrate how such solutions may be constructed, let us consider the 
optimal realization of the 
${\bf \Phi}$--field,   
${\bf \Phi}(\tau) = 2\pi {\bf W}\tau/\beta$. 
Employing Eq.~(\ref{bc2}), one finds 
\begin{equation}
{\bf Q}(x=0) =  
\delta^{ab}
\mbox{sign}(\varepsilon_{m-W_a}) \delta_{mk}\, .
                                            \label{bc3}
\end{equation}
This is equivalent to the local, replica dependent shift of the chemical
potential on $W_a$ Matsubara units.  
Consider e.g. the simplest non--trivial winding number  configuration 
consistent with Eq.~(\ref{rest}) having only two non--zero components of 
${\bf W}$:
\begin{equation}  
W_a = 1;\,\, a\in [1,n]\,; \,\,\,\,\, \,\,  W_{b} = - 1;\,\, b\in [n+1,2n]\, 
                                           \label{inst}
\end{equation}
and $W_c=0$ for $c\neq a,b$. 
In this case one may satisfy the boundary conditions 
Eqs.~(\ref{bc1}), (\ref{bc3}) along with ${\bf Q}^2=1$ by choosing 
\begin{equation}
\left(
\begin{array}{rr}
Q^{aa}_{{\pi\over\beta}{\pi\over\beta} } &
Q^{ab}_{{\pi\over\beta}-{\pi\over\beta} } \\
Q^{ba}_{-{\pi\over\beta}{\pi\over\beta} } &
Q^{bb}_{-{\pi\over\beta}-{\pi\over\beta} } 
\end{array}
\right) = 
\left(
\begin{array}{cc}
\cos \theta(x) & \sin \theta(x) \\
\sin \theta(x) & -\cos\theta(x)
\end{array}
\right)\, ,
                                           \label{o2}
\end{equation}
where $\theta(0) = \pi$, $\theta(L) = 0$ and all other 
elements of ${\bf Q}$ are equal to those of ${\bf \Lambda}$. 
The Usadel equation takes the form of the anharmonic pendulum equation  
\begin{equation} 
L_T^2 \nabla^2 \theta - \sin \theta = 0\, .  
                                         \label{usad1}
\end{equation} 
Its solution may be written in terms of the elliptic integrals.  
The corresponding action is equal to 
\begin{equation} 
S_{w} = 
G{L\over 2 L_T}\left[
{4\over \sqrt{\kappa} }E(\kappa) - 
{1-\kappa \over \kappa} {2L \over L_T}
\right] + O(n)
 \, ,
                                      \label{act}
\end{equation}
where $\sqrt{\kappa}K(\kappa)=L/L_T$ and $K(\kappa)$ and $E(\kappa)$ are 
the complete elliptic
integrals of the first and second kind. In the two limiting cases one
obtains (in the $n\to 0$ limit)
\begin{equation} 
S_{w} = 
\left\{
\begin{array}{ll}
{\pi^2\over 4}G\left(1 +{4 T\over \pi E_T} \right) \,\,\,\,& T\ll E_T\, ;\\
2 G(L_T),    &            T\gg E_T \, , 
\end{array}
\right.
                                         \label{act1}
\end{equation}
where $G(L_T) \equiv e^2\nu D/L_T=G\sqrt{2\pi T / E_T} $ 
is the dimensionless conductance of the length $L_T$ of the wire.
In the zero temperature case the instanton spreads uniformly over the 
wire, 
$\theta(x) = \pi (L-x)/L$. 
In the opposite limit, $L\gg L_T$, it is confined to the region of the size 
$L_T$ near the dot: 
$\theta(x) = 2\pi -4\arctan[\exp\{x/L_T\}]$.   
The low temperature (short  
wire) limit may be directly compared to  Nazarov's result, 
Eq.~(\ref{Naz1}) (we calculate 
$\langle \tilde N^2 \rangle$ rather than 
$\langle |\tilde N| \rangle$ thus the factor of two difference). 
In general case Eq.~(\ref{act}) provides the 
entire temperature and wire-length dependence of the 
oscillation amplitude (with the exponential accuracy). 
Winding number configurations other than Eq.~(\ref{inst}) are exponentially 
less probable. 
The analytical continuation, $n\to 0$, is straightforward if one notices
that there are precisely $n^2$ different 
configurations like Eq.~(\ref{inst}) with the same action and
$q-q'$ dependence. 
Thus the  $n^2$ factor  in  Eq.~(\ref{corr}) is cancelled and the remained
expression is $n$-independent. 
It is important to notice that the rotational symmetry between the replicas 
is broken down to the permutation one by the choice of the winding number 
configuration. As a result
there is  no degenerate saddle point manifold, which 
$n$-dependent volume plays a central role in the case of non--interacting 
level statistics   \cite{Kamenev99a}.       

One can define the current operator on the instanton trajectory 
as ${\bf J} \equiv i{\bf Q}\nabla {\bf Q} = \sigma_y \nabla \theta(x)$, where 
$\sigma_y$ is the Pauli matrix in the space defined in Eq.~(\ref{o2}). 
In the limit of the long wire, $L\gg L_T$, the current is non--uniform along 
the wire since, $\nabla \theta\neq {\mbox const}$. By virtue of the continuity 
equation,  this seemingly leads to the charge accumulation in the region 
$x\sim L_T$. Where this  indeed the case, our model of the 
non--interacting wire would loose its validity and finite temperature results 
should be modified. This is not the case, since the current, 
${\bf J}$, is pure replica off--diagonal (because of the $\sigma_y$ matrix)
and therefore completely decouples from the Coulomb interactions. 
Interpretation of this counter--intuitive phenomena is that the replica 
structure of the instanton takes care of the mesoscopic fluctuations effect.
It describes sample specific energy (and thus temperature) 
dependence of the transmission coefficients, introducing retardation 
neglected in  Nazarov  approach. This retardation is not 
associated with the accumulation of any real charge in some portion 
of the wire.

%
%
%

Finally, we address the 2D setup with the dot  of the radius $d$ 
placed at the center of a 2D disordered disk of the radius $L$. 
In this case the elementary instanton is given by 
Eqs.~(\ref{inst}), (\ref{o2}) with 
$\theta=\theta(r)$ which is a solution of the 
following equation 
\begin{equation} 
L_T^2 \partial_r (r \partial_r\theta) - 
r \sin \theta = 0
                                                          \label{2D}
\end{equation}
with $\theta(d)=\pi$ and $\theta(L)=0$; $r$ is the polar radius.
For $L\ll L_T$ the equation is solved by \cite{foot6}  
\begin{equation} 
\theta(r) = \pi\left[ 1-{\ln r/d \over \ln L/d}\right]\, ; 
                                                          \label{2Da}
\end{equation}
with the corresponding action 
\begin{equation}
S_{w} = {\pi^2\over 4}\, {2\pi e^2\nu D\over \ln L/d}  \, . 
                                                          \label{act2}
\end{equation}
At higher temperature, $d\ll L_T < L$, with the logarithmic accuracy 
one must substitute $L_T$ instead of $L$ in Eq. (\ref{act2}). 
By analogy with Eq.~(\ref{act1}) the final result may be written as 
$S_{w}=\pi^2 G(L_T)/4$, where $G(L_T)=2\pi e^2\nu D/ \ln L_T/d$ is 
the conductance of 2D disk with the outer radius $L_T$ and the 
inner one $d$. 
Note that the temperature dependence of the 2D action is 
$[\ln D/(d^2 T)]^{-1}$ much slower than $\sqrt{T}$ in one dimension.

In conclusion, we have shown  that the residual charge quantization in the
diffusive interacting systems may be described 
by  the instantons of the interacting 
NL$\sigma$M. These instantons are replica non--symmetric finite action
configurations of the ${\bf Q}$--field, 
which extremize the action with the "twisted" boundary conditions. As an
illustration, we have calculated the 
(exponentially small) amplitude of the Coulomb blockade oscillation in the
open quantum dot 
connected to 1D (2D) diffusive conductor of arbitrary length (radius). 
Although the instantons are exponentially rare in the diffusive regime, they are 
responsible for the  
charge quantization and inhomogeneity -- the features which dominate the
physics of the insulating phase.  One may expect them to be increasingly 
important as the insulator regime is approached.

Discussions with L. Levitov and A. Shytov have initiated this project. 
Valuable comments  of  I. Aleiner, A. Andreev, M. Feigelman, Y. Gefen and   
I. Gruzberg  are highly acknowledged. 
This research was supported in part by the BSF grant N  9800338.

\ecols

\begin{thebibliography}{10}
\bibitem{foot1} To disregard the thermal smearing of the Coulomb blockade,
the inequality $T\ll E_c$ (see Eq.~(\ref{Ham})) is assumed hereafter.
\bibitem{foot4}The better quantity to average is $\Re\ln \tilde N(q)$, 
\cite{Nazarov99}. It may be calculated in the same way with a slightly more 
complicated algebra.
\end{thebibliography}
\end{document}